\documentclass[12pt]{iopart}

\usepackage{graphicx} 
\begin{document}

\title[]{Common Organizing Mechanisms in Ecological and Socio-economic Networks}

\author{Serguei Saavedra$^{1,2,3,4}$, Felix Reed-Tsochas$^{3,5}$ and Brian Uzzi$^{1,2}$}

\address{$^1$ Kellogg School of Management, Management and Organizations Department, Northwestern University, Evanston, Illinois, USA, 60208

$^2$ Northwestern Institute on Complex Systems, Northwestern University, Evanston, Illinois, USA, 60208

$^3$ CABDyN Complexity Centre, Oxford University, OX1 1HP, UK

$^4$ Oxford University Centre for Corporate Reputation, Said Business School, Oxford, OX1 1HP, UK

$^5$ James Martin Institute, Said Business School, Oxford University, UK, OX1 1HP
}
\ead{\mailto{s-saavedra@kellogg.northwestern.edu}, \mailto{felix.reed-tsochas@sbs.ox.ac.uk}, \mailto{uzzi@kellogg.northwestern.edu}}

\begin{abstract}
Previous work has shown that species interacting in an ecosystem and actors transacting in an economic context may have notable similarities in behavior. However, the specific mechanism that may underlie similarities in nature and human systems has not been analyzed. Building on stochastic food-web models, we propose a parsimonious bipartite-cooperation model that reproduces the key features of mutualistic networks - degree distribution, nestedness and modularity -- for both ecological networks and socio-economic networks. Our analysis uses two diverse networks. Mutually-beneficial interactions between plants and their pollinators, and cooperative economic exchanges between designers and their contractors. We find that these mutualistic networks share a key hierarchical ordering of their members, along with an exponential constraint in the number and type of partners they can cooperate with. We use our model to show that slight changes in the interaction constraints can produce either extremely nested or random structures, revealing that these constraints play a key role in the evolution of mutualistic networks. This could also encourage a new systematic approach to study the functional and structural properties of networks. The surprising correspondence across mutualistic networks suggests their broadly representativeness and their potential role in the productive organization of exchange systems, both ecological and social.
\end{abstract}

\maketitle

\section{Introduction}
The analogy between ecological and economic systems is not new. Biologists have always being intrigued about the economic aspects of nature \cite{Ricklefs,Tilman}, and economists and sociologists have taken insight from biological systems to shed new light on the factors shaping socio-economic systems. For instance, researchers have adapted biological models that focus on constructs such as niche width, resource partitioning, or specialization and generalization to explain the birth and death rates
of organizational populations \cite{Hannan,Carroll}. More recently, multidisciplinary approaches have led to the discovery of significant structural similarities across different network domains, including biological and socio-economic networks \cite{Newman1,Dorogov,Caldarelli}. This has awakened an even more spirited search for common structural properties between ecological and economic networks \cite{Sugihara,Martinez}, and pointed to a greater need for work on mutually-beneficial interactions across realms. 

Cooperation \cite{Hamilton,Axelrod,Hammerstein} is a central concept in biological and social studies, and although the evolution of cooperation was initially modeled for homogeneous populations, subsequent work has also included spatial effects \cite{Nowak}. The recent development of simulation models for evolutionary games on graphs \cite{Ohtsuki} and collaborative social networks \cite{Guimera1} provides a starting point for addressing the question of how cooperative structures are assembled. In ecology, contemporary research on mutualistic networks provides us with an increasingly detailed picture of the complex set of cooperative interactions between different species in an ecosystem, and demonstrates that purely local interactions can generate highly structured macroscopic patterns of mutually beneficial exchanges \cite{Bascompte1}. 

However, despite the increasing interest in cooperative systems, we currently lack models that allow us to connect the cooperative behavior at the level of individuals with emergent global network properties. Furthermore, although cooperation appears as distinctive characteristic at different levels of organization ranging from groups of animals to human society \cite{Nowak2}, it has been difficult to find empirical evidence showing similar patterns of cooperation shared across these levels. In previous work \cite{Saavedra2}, we have proposed a bipartite-cooperation model (BC model) that can replicate key properties of mutualistic networks. To test the BC model across ecological and socio-economic networks, we have used ten large pollination datasets that have been compiled in the literature, and a unique and extensive, economy -wide dataset of designers and contractors engaged in joint production in the New York City garment industry. In the present paper, we attempt to go beyond showing associations in the assembling principles of these ecological and socio-economic networks to show the effects of different organizing mechanisms on the hierarchical arrangement of these networks. In Sections 2 and 3 we discuss some of the main features of mutually-beneficial interactions in ecological and socio-economic networks respectively. Sections 4, 5 and 6 describe the BC model, the empirical data and the validation of the BC model respectively. Section 7 introduces a further justification of the BC model using empirical data. In Section 8 we use the BC model to study the effects of different organizing mechanisms and interaction constraints on the hierarchical arrangement of empirical networks. Finally, Section 9 summarizes our conclusions and overall findings.

\section{Ecological networks}
In ecology, mutualistic networks are formed by the mutually-beneficial interactions between populations of different species (e.g. P for plants and A for animals) \cite{Bascompte1}. Species in class P offer rewards with certain characteristics to attract species in class A. These individual attributes, determined by their own reward traits, may also have evolved to reduce exploitation and favor mutualism \cite{Bronstein}. Species in class A foraging for resources can benefit from the rewards offered by a given species in class P if the respective foraging traits (e.g. efficiency, morphology, behavior) and reward traits (e.g. quantity, quality, availability) are complementary \cite{Jordano,Waser}. External factors such as the environmental context (e.g. population density, geographic variation) attenuate or amplify the value of reward and foraging traits, and impact the number of potential partners that a given species cooperates with \cite{Noe,Olesen,Guimaraes2}. Furthermore, Rezende et al.\cite{Rezende} have shown that mutualistic networks exhibit hierarchical constraints introduced by phylogenetic relationships between species in the same class, which impact mutualistic interaction patterns by favoring ecological similarity.

Recent work has found key structural properties in mutualistic networks. Mutually-beneficial interactions between species exhibit broad-scale distributions and a significant presence of asymmetric interactions (i.e. links connecting high-degree to low-degree nodes), which can be the result of mechanisms such as aging, forbidden interactions, capacity constraints, etc. \cite{Jordano,Amaral,Bascompte3}. Mutualistic networks also display nested and modular structures that differ from appropriate random assemblages, which play a crucial role in structural robustness and function of these networks \cite{Bascompte2,Olesen}. In addition, studies have found that the relationship between the number of species $S$ and mutually-beneficial interactions $L$ follows a power-law given by $L=S^{\alpha}$, where $\alpha=1.13$ \cite{Bascompte2}. This suggests that mutualistic networks might display a universal mode of organization that appears to enhance the ability of species to respond to environmental changes and competition pressures \cite{Bascompte1,Thompson,Bastolla}.

\section{Socio-economic networks}
Exchanges and beneficial relationships between members in social and economic contexts are also known to give rise to large-scale cooperative networks through which resources and information can flow \cite{New,Uzzi,Saavedra3,Saavedra1}. In an organizational context, firms are characterized by a set of reward or organizational traits (e.g. firm size, competitive niche space, brand positioning), which are modulated by hierarachical barriers generated by differences in status, which limit the number and range of potential partners \cite{Carroll,Hannan,Podolny,Gould}. Hence, cooperation between two different classes of firms (e.g designers and contractors in a manufacturing industry) can also be subject to structural constraints equivalent to those found in ecological networks. These constraints depend on the traits of firms and the complementarity between traits of potential partners, as well as hierarchical relationships between firms in the same class. Note that the values associated with reward traits and interaction barriers between firms are not absolute, but modulated by the specific market context that the firms operate in \cite{Carroll,Podolny}.

\begin{figure}[t]
\begin{center}
\begin{tabular}{c}
\includegraphics[width=3.5in]{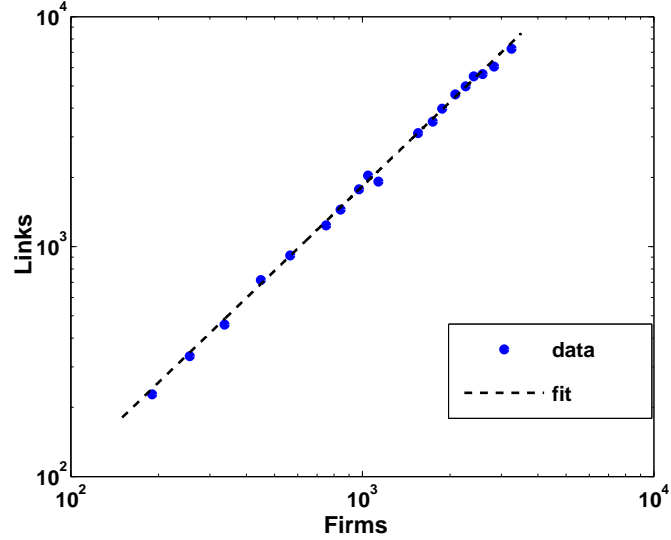}
\end{tabular}
\end{center}
\caption{Scale invariability. The figure shows the relationship on a log-log scale between the total number of firms $F$ and links $L$ in the NYGI network from 1985 to 2003 (dots). The solid line is the fit to the data defined by $L=F^\alpha$ with $\alpha=1.22\pm0.01$ ($R^2=0.98$). This reveals that although the NYGI network experienced a dramatic contraction over the years in its total number of links and firms \cite{Saavedra1}, declining from over 3000 firms in 1985 to 190 firms in 2003, the proportion between firms and links remained constant.}
\label{fig1}
\end{figure}

Interestingly, in common with mutualistic networks, studies have found that the number of collaborative partners in different socio-economic networks follows a broad-scale distribution characterized by asymmetric interactions, which not only prevent the network from collapsing, but also enhance the efficiency of the network \cite{Saavedra1,May}. Previously, we found that although the total number of firms $F$ in socio-economic network varies from year to year, the relation between the total number of firms and links $L$ also follows a constant relation \cite{Saavedra2}. Figure \ref{fig1} shows that this relation is defined by a power-law $L=F^{\alpha}$ with $\alpha=1.22$. In addition, structurally cohesive social networks do enhance a hierarchical nesting of groups \cite{Moody}. These results suggest that mutualistic networks and cooperative socio-economic networks might have both structural and organizing mechanism in common.   

\section{The bipartite-cooperation model}
In theoretical ecology, stochastic models incorporating simple interaction rules have recently proved remarkably successful at reproducing overall structural properties of real food webs \cite{Williams,Cattin}. Food webs are formed by interactions between predators and preys that reflect all possible flows of energy and biomass between species in an ecosystem. It has been shown that these models need to satisfy only two basic conditions about the distribution of niche values and feeding ranges to reproduce many aspects of the complex network of predator-prey interactions, using species richness $S$ and connectance $L/S^2$ as the only input parameters \cite{Stouffer}. These two basic conditions are: (1) an ordered set of species' niche values; (2) an exponentially decaying probability of preying on species with lower niche values. Note that niche values can be better explained by a hierarchical ordering of body-sizes \cite{Otto}. Recently, more detailed models have incorporated higher levels of constraints in order to reproduce the actual links observed in real food webs \cite{Petchey,Allesina,Allesina2}. Hence, the notable success in modeling the complex set of predator-prey interactions between species has suggested that simple models may equally well account for the structure of networks generated by cooperative partner-partner interactions between two distinct classes of species, as in the case of plant-animal mutualistic networks \cite{Guimaraes,Santamaria}.

Here, building on food web models \cite{Williams,Cattin}, we developed the BC model that applies when members in a network can be divided into two distinct classes (e.g. class $P$ and class $A$). Hence, the inputs for the BC model are the size of class $A$, the size of class $P$ and the total number of links $L$, all of which are given directly by the empirical data. The BC model consists of a specialization and an interaction mechanism, which are motivated by the fact that specialists (members with low number of partners) will face less competition and fewer interaction barriers by cooperating with generalists (members with high number of partners) \cite{Bastolla}.

\begin{enumerate}
\item \textbf{Specialization}. The specialization rule determines how many partners $l_p$ each member $p$ $\in$ $P$ will cooperate with, and is calculated following the nested-hierarchy model \cite{Cattin}. The number of partners $l_p$ is given by $l_p=1+Round(\frac{(L-|P|)t_{Rp}\lambda_p)}{\sum_j t_{Rj}\lambda_j})$, where $|\cdot|$ denotes set cardinality and $Round(\cdot)$ is the nearest integer function. Here, $t_{Rp}$ is the reward trait associated with member $p$, uniformly drawn from $[0,1]$, attenuated or amplified by an external constraint $\lambda_p$, randomly drawn from an exponential distribution (described below), that accounts for effects such as geographic variation and population diversity. Reward traits $t_{Rp}$ are the result of a hierarchical process, which in the BC model corresponds to the generation of an ordered sequence in trait space, so $t_{Rp}$ plays an equivalent role to the niche value in the niche model \cite{Williams}.

\item \textbf{Interaction}. The interaction rule determines which members $a$ $\in$ $A$ cooperate with each member $p$ $\in$ $P$. Mutualistic interactions are hierarchically limited by the complementarity between reward traits $t_{Rp}$ for $p$ $\in$ $P$ and foraging traits $t_{Fa}$ for $a$ $\in$ $A$, which both are uniformly drawn from $[0,1]$. Members from class $P$ are sorted according to their reward trait $t_{Rp}$ in ascending order given their ability to attract partners; whereas members from class $A$ are sorted according to their foraging traits $t_{Fa}$ in descending order given their ability to cooperate with partners. Starting from the first -specialist- member $p_i$ and continuing sequentially subject to $t_{Rp_i}>\lambda_{l_{pi}}$, each link $l_{pi}$ is connected to the first -generalist- member $a' \in A'$, where $A'$ is the subset of members in $A$ that have not already been linked to by another member $p\neq p_i$. If $t_{Rp_i}\leq \lambda_{l_{pi}}$, the link is randomly connected to another member $a'' \in A''$ that belongs to a previously randomly selected member $p_j$ with lower trait value, where $A''$ is the subset of members in $A$ that have been allocated links in a previous timestep. This is to say, for each individual link of the new member $p_i$, first we randomly select a different member $p_j$ with lower trait value, and second we randomly select a member $a''$ from the partners of $p_j$. This process is repeated independently for each individual link. Note that $\lambda_{l_{pi}}$ corresponds to an external constraint associated with each link $l_{p_i}$ that accounts for interaction barriers such as competition and population density. This is drawn randomly from the same exponential distribution as $\lambda_p$ (described below). In applying the interaction rule, if the supply of partners in either subset $A'$ or $A''$ is exhausted before all $l_{pi}$ links have been allocated, then partners in the other subset are chosen to instead. The effect of this additional rule is equivalent to imposing a size limit on the population \cite{Guimaraes}. The BC model is initialized by connecting the first member $p_i$ to $l_{pi}$ members in $A'$.

\end{enumerate}

In line with prior food web models \cite{Stouffer}, and in order to account for inhomogeneous effects in the population, we assume that external constraints $\lambda_{p}$ and $\lambda_{lp}$ are randomly drawn from an exponential distribution given by $p(x)=\beta exp(-\beta x)$, with $\beta =|P|(|A|-1)/(2(L-|P|))-1$. Note that $L/(A\cdot P)$ is equivalent to the connectance of a bipartite network. As has been shown \cite{Stouffer}, this exponential distribution is equivalent to a beta distribution under low-connectance levels. We also found that the BC model produces similar results using a beta distribution of the form $p(x)=\beta (1-x)^{\beta-1}$  (see supplementary information in \cite{Saavedra2}).

\section{Empirical data}
In the present paper, we use a diverse set of ten extensive plant-animal pollination networks compiled in the literature (see Table), which can clearly be distinguished from random assemblages \cite{Bascompte2}. Mutually-beneficial interactions in a pollination network are formed when an animal such as a bee or wasp gets food in form of nectar in exchange of transferring pollen from one plant to another of the same species. Hence, mutually-beneficial interactions are established by the transportation mechanism, which enables fertilization and sexual reproduction for plants.

Our observed socio-economic network uses information collected by the workers union UNITE (Union of Needle Trades and Industrial and Textile Employees) on the designer-contractor network of the New York Garment Industry (NYGI). UNITE has organized around 90 percent of all firms in the NYGI, and has developed a reliable system to ensure the validity of transaction data \cite{Uzzi}. The dataset includes approximately 700,000 designer-contractor bilateral exchanges from January 1985 to December
2003. We formed yearly network snapshots from 1985-2003 given the seasonality and volatility of the NYGI industry \cite{Saavedra1}. A cooperative interaction exists between a designer and contractor if they co-produce a garment in a particular year. For example, the typical production process in a year begins with a designer that develops a line of clothing. Each garment in the line is made into a sample prototype, which is disassembled into its component parts such as shelves, collars, waistbands, and so forth. The components of the sample are then sent by the designer to contractors that cut components from fabric in lots large enough to be mass produced. The cut fabric is then sent by the designer to sewing contractors that sew the fabric together into the garments that are sold directly to consumers at retailers. All firms are free to make connections of their own choice; there is no governing body that suggests or mandates connections (see \cite{Saavedra1} for more details on the data).

\section{Model validation}
To analyze the performance of the BC model, we measure the ability of the BC model at reproducing the degree distribution, nestedness and modularity of both pollination networks and the NYGI networks. These network metrics are key organizational features present in networks formed by bipartite cooperation \cite{Jordano,Bascompte2,Olesen,Bastolla}. To capture the statistical relevance of our model-generated networks, we use a Kolmogorov-Smirnov (KS) comparison test and a $z$-score analysis (i.e. normalized errors) to test the overall goodness of fit of the BC model to the empirical data. All comparisons are based on 1000 model simulations for each network (for comparisons of the BC model to other mutualistic models see supplementary information in \cite{Saavedra2}). In the BC model, plants and designers are treated as members of class $P$, and animals and contractors as members of class $A$. This also follows the rationale that animals and contractors might experience higher competitive pressures than animals and contractors, given the differences in population size. From the Table we can see that the ratio between plants animals is $P/A<0.5$, which also plays a significant role limiting the appearance of scale-free distributions \cite{Guimaraes}.

\subsection{Degree distribution}
\begin{figure}[t]
\begin{center}
\begin{tabular}{cc}
\includegraphics[width=3.5in]{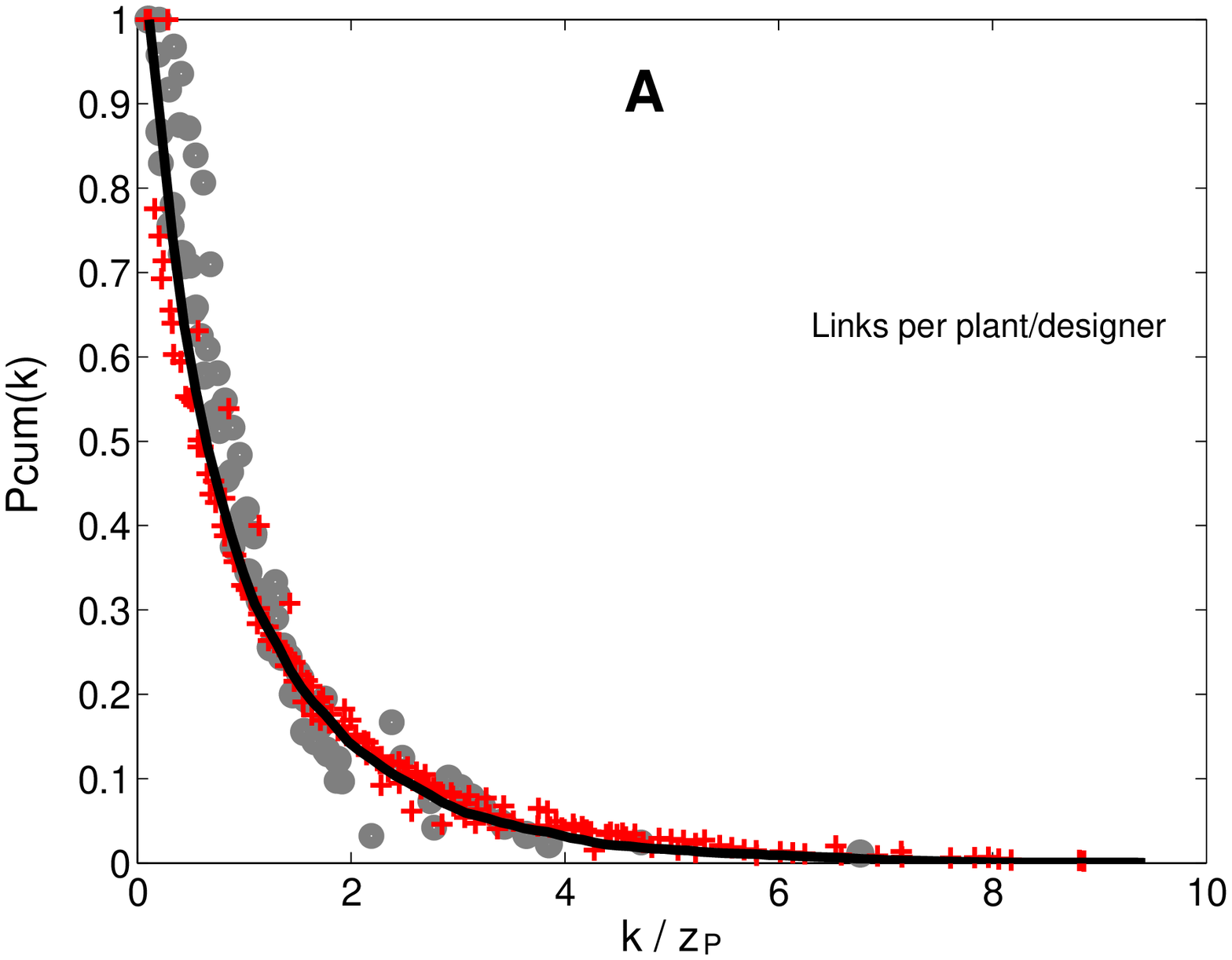}&\includegraphics[width=3.5in]{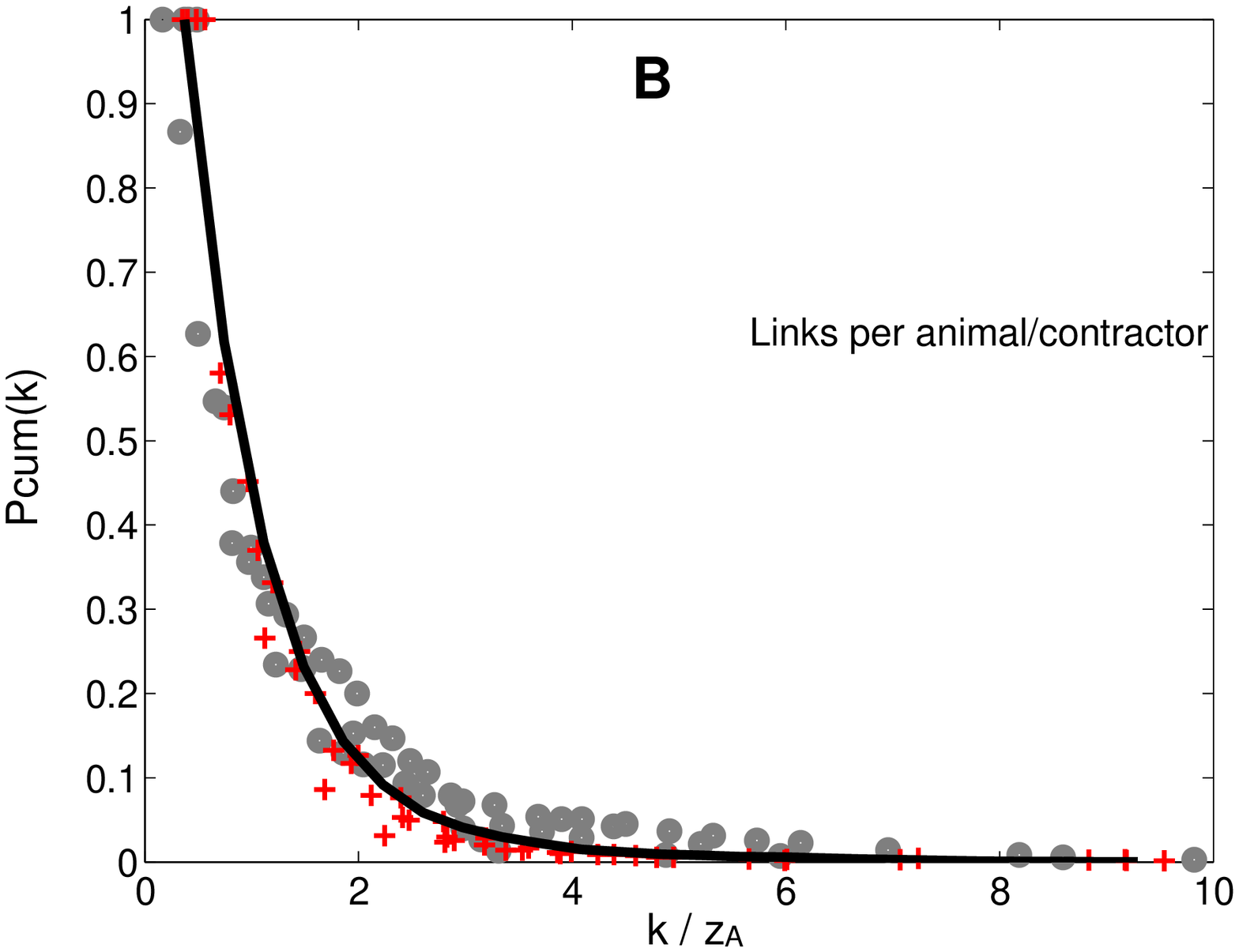}
\end{tabular}
\end{center}
\caption{Cumulative degree distribution. Panel \textbf{A} and Panel \textbf{B} show the scaled cumulative degree distribution $P_{cum}(k)$ for members of class P (plants, designers), and members of class A (animals, contractors) respectively. The number of partners $k$ is scaled by a multiplicative factor of $1/z_P$ for members of P, and $1/z_A$ for members of A, where $z_P=L/P$ and $z_A=L/A$. Solid symbols correspond to pollination networks and crosses correspond to NYGI networks. Note that all distributions collapse into a single curve. The solid line corresponds to the model-generated distributions averaged over 1000 simulations.}
\label{fig2}
\end{figure}

The degree distribution $P(k)$ is a widely used statistical metric that measures the probability that a node has up to $k$ network connections \cite{Newman1,Dorogov,Caldarelli}. This measure affects network growth and decline in different complex networks \cite{Saavedra1,Albert}. Figures \ref{fig2}A and \ref{fig2}B show the scaled cumulative distribution for members of class P and members of class A respectively. Note that pollination networks (solid symbols) and NYGI networks (crosses) exhibit the same patterns in both degree distributions. The solid line corresponds to the model-generated degree distributions. The Table shows that the BC model reproduced most of the empirical distributions.

\subsection{Nestedness}
Nestedness is a concept applied in ecology to metacommunity populations, where sites with low biodiversity constitute proper subsets of sites with higher biodiversity \cite{Atmar}. This has been extended to ecological networks, where a network is said to be nested when specialist species interact with proper subsets of the ecological interactions of generalist species \cite{Bascompte3}. To calculate nestedness $N$, we use the BINMATNEST program \cite{Rodriguez}. This program uses a genetic algorithm to find the matrix configuration that minimizes the level of disorder, and calculates an isocline curve that is used to measure the distance to the situation of perfect order or nestedness for each pair-wise interaction. Here nestedness is defined in the interval [0,1], where 1 corresponds to a perfectly nested network. In addition, we use a null model (i.e. randomized version) to compare whether the empirical networks and model-generated networks are indeed different from random assemblages. For this, we use the null model II proposed in \cite{Bascompte3}, which has been shown to produce conservative results with respect to nestedness \cite{Ulrich}.

\begin{figure}[t]
\begin{center}
\begin{tabular}{cc}
\includegraphics[width=3.5in]{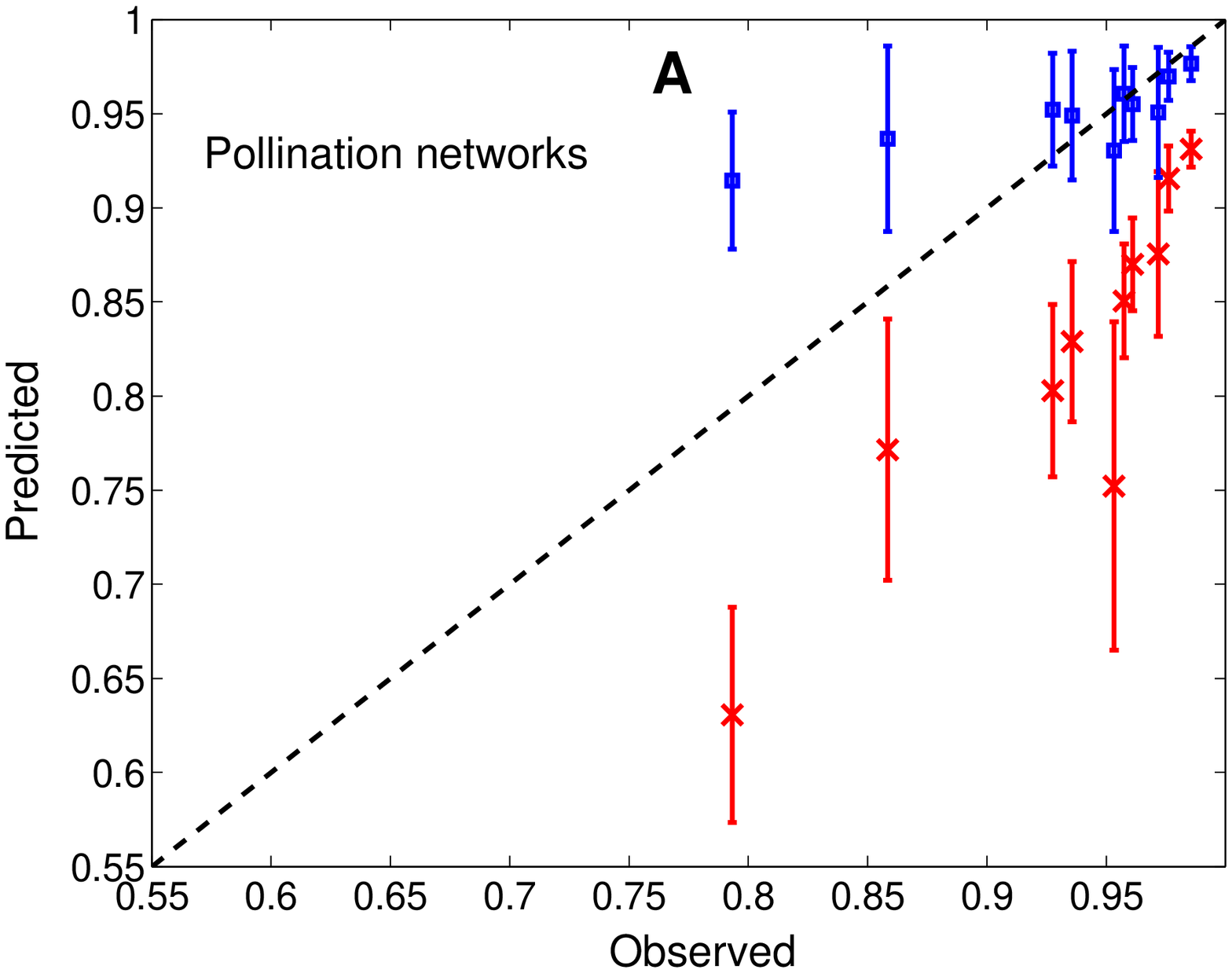}&\includegraphics[width=3.5in]{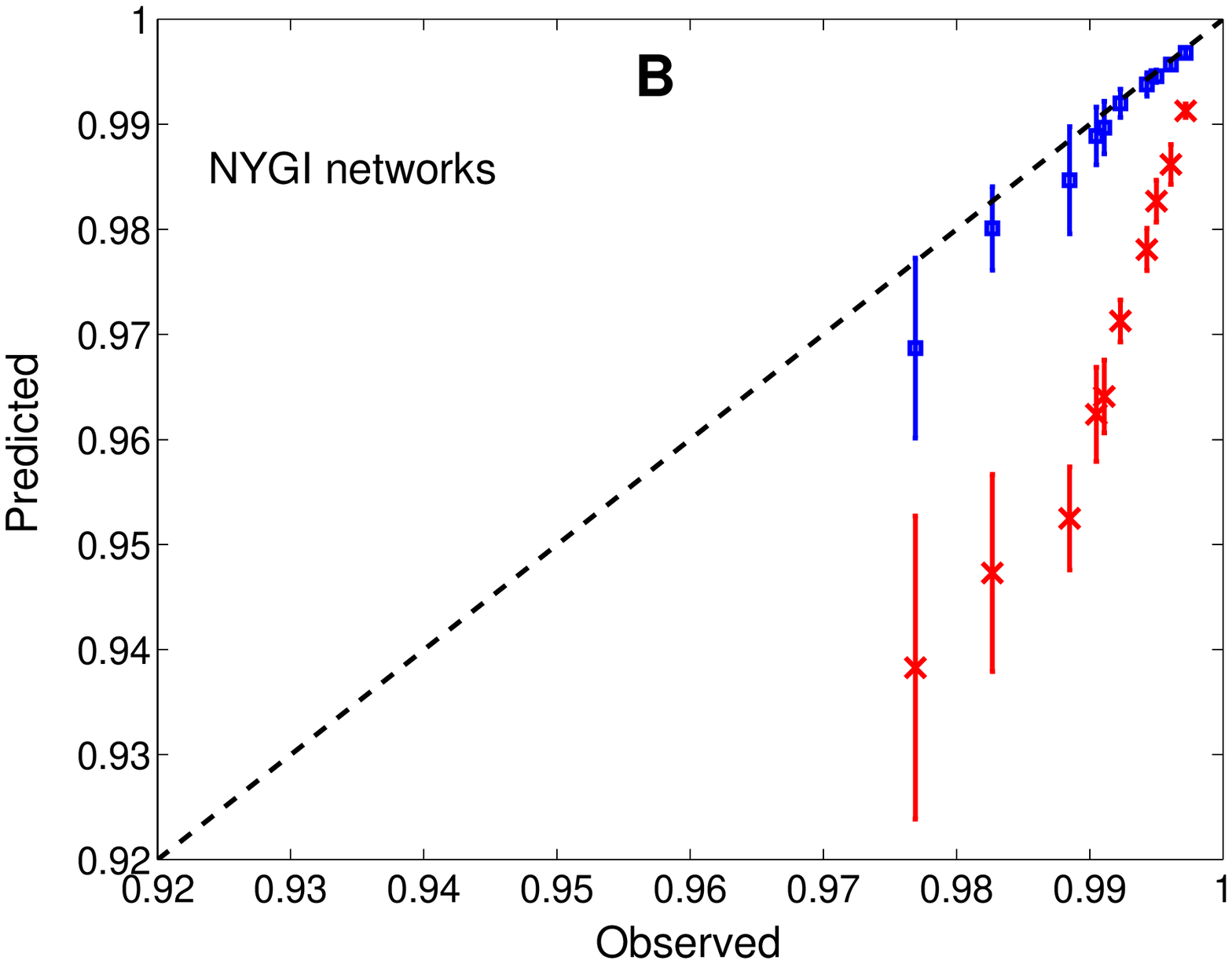}
\end{tabular}
\end{center}
\caption{Nestedness. Panel \textbf{A} and Panel \textbf{B} compare the random-generated (red) and model-generated (blue) nestedness values to the observed nestedness values for the pollination and NYGI networks respectively. The dashed line would correspond to a perfect agreement with the observed values. Bars are equivalent to $avg.\pm2s.d.$ values. Here, a nestedness of 1 means a perfectly nested matrix.}
\label{fig3}
\end{figure}

Figure \ref{fig3}A and \ref{fig3}B displays the nestedness pattern for empirical networks (dashed line), random assemblages (red), and model-generated networks (blue) for pollination networks and NYGI networks respectively. The dashed line would correspond to a perfect agreement with the observed values. Bars are equivalent to $avg.\pm2s.d.$ values -a nestedness of 1 means a perfectly nested matrix-. Note that the BC model always performs significantly better than random assemblages. To capture the statistical relevance of our model-generated nestedness values, we use a $z$-score analysis given by $Z=(N-\overline{N}_m)/\sigma_m$ where $N$ is the nestedness value for the empirical network, and $\overline{N}_m$ and $\sigma_m$ are the average nestedness value and the standard deviation for the model-generated network respectively. Our Table shows the high goodness of fit of the BC model at reproducing the empirical nestedness values for the observed ecological and NYGI networks. Different nestedness algorithms do not change the qualitative nature of our results \cite{GuimaraesEM,Almeida}.

\subsection{Network modularity}
The third characteristic that we analyze is network modularity \cite{Girvan}. The modularity of a network depends on its number of groups or modules and on the deviations from the number of groups expected in a properly randomized network \cite{Girvan}. The modularity values $Q$ for the networks were calculated using the one-mode optimization algorithm \cite{Guimera2}. A good partition generates many within-community links and as few as possible between-community links. This algorithm does not discriminate between plants and pollinators or designers and contractors, where communities comprise nodes from both classes. This is important because we want to extract only cooperative units from the network, and using a two-module partition algorithm \cite{GuimeraB} (i.e. modules with only one class of nodes), we could erroneously form communities between competitive members instead. 

\begin{figure}[t]
\begin{center}
\begin{tabular}{cc}
\includegraphics[width=3.5in]{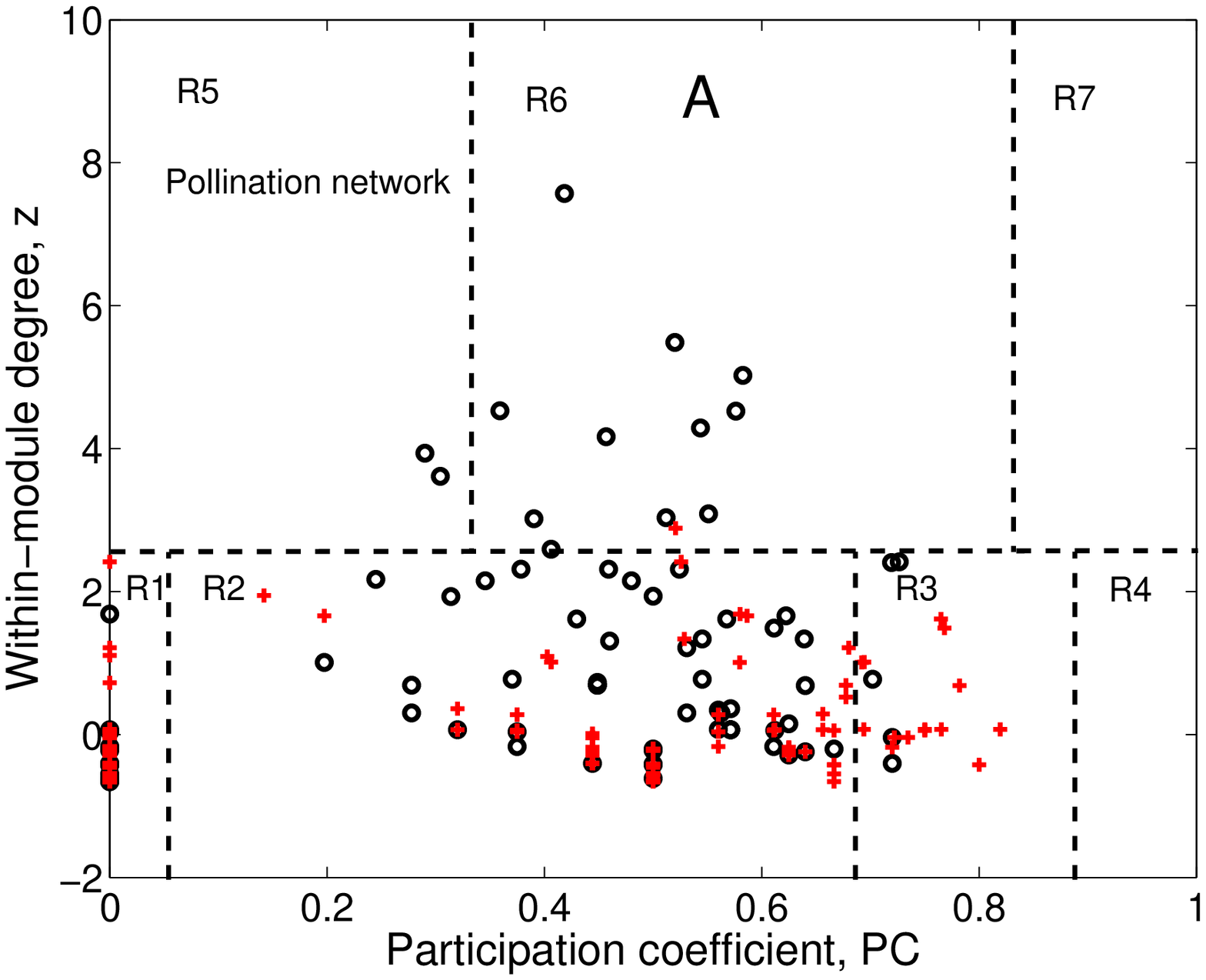}&\includegraphics[width=3.5in]{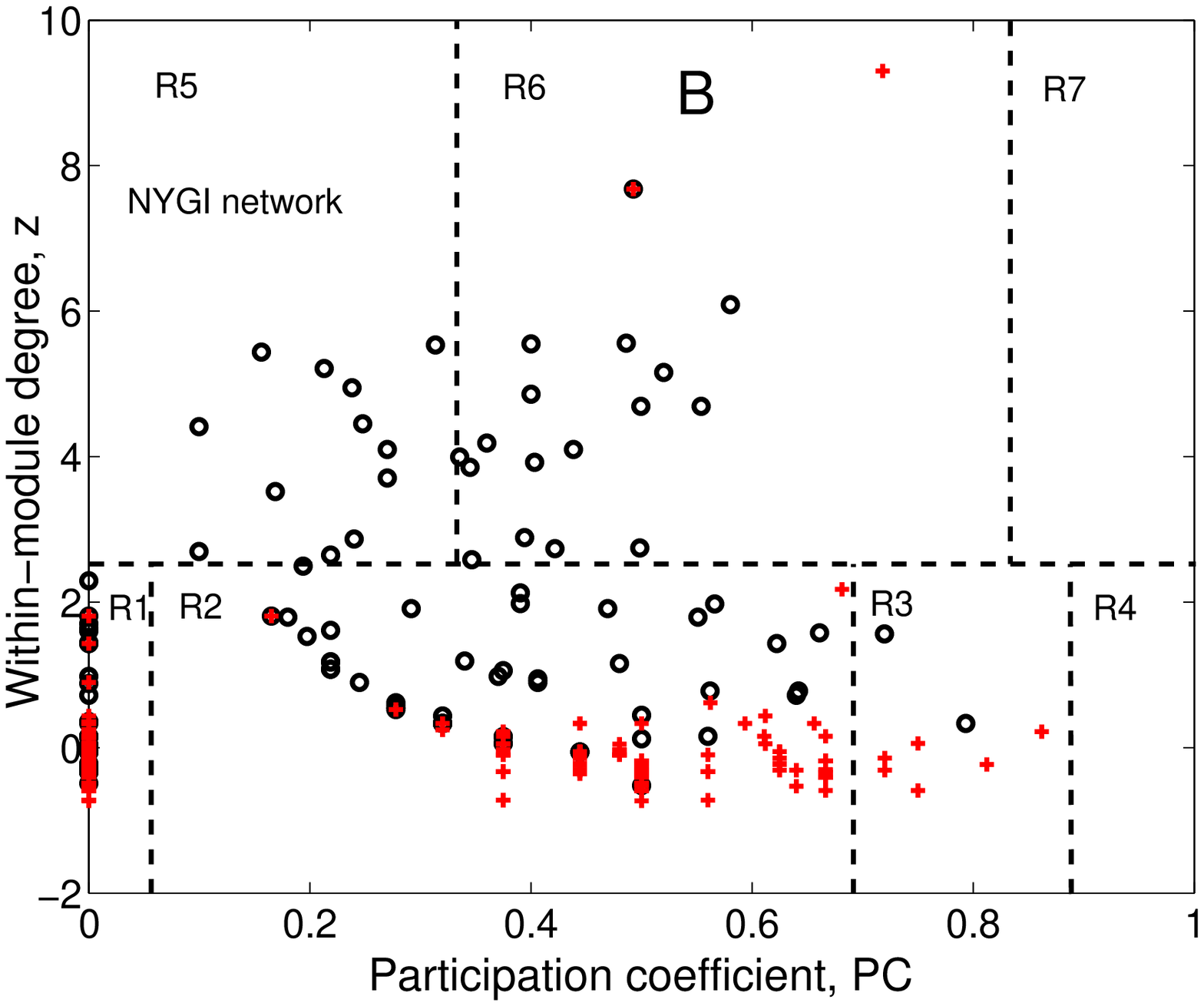}
\end{tabular}
\end{center}
\caption{Empirical connectivity roles. Panel \textbf{A} and Panel \textbf{B} show the connectivity-role space for members of the Kato and Miura (1996) network and the 1997 NYGI network respectively. Plants and designers are shown in black dots, whereas animals and contractors in red crosses. Note that the two networks present similar patterns. The classification is as follows \cite{Guimera2}: Nodes with $z\geq2.5$ are classified as module hubs and nodes with $z<2.5$ as non-hub nodes. In addition, hubs and non-hub nodes are classified according to the corresponding participation coefficient $PC$. For non-hub nodes (R1) is for nodes with all their links connected within their own module $P\leq0.05$, (R2) is for nodes with most of their links connected within their own module $0.05<P\leq0.62$, (R3) is for nodes with many links connected to other modules $0.62<P\leq0.80$ and (R4) is for nodes with their links homogeneously connected to all other modules $P>0.80$. For hub-nodes, (R5) is for nodes with most of their links connected to their own module $P\leq0.30$, (R6) is for nodes with most of their links connected to other modules $0.30<P\leq0.75$ and (R7) is for nodes with their links homogeneously connected to all other modules $P>0.75$.}
\label{fig4}
\end{figure}

To test our fit, we are not looking for a specific partition of the network \cite{Fortunato}, but a standard and well defined benchmark of comparison. Recent work has shown that half of the pollination networks studied in the present paper are modular (see Table) \cite{Olesen}. Using again a $z$-score analysis, the Table shows a surprisingly high degree of correspondence between the five empirical modularity values and the model-generated ones. For all the observed NYGI networks, we always found higher modularity values compared to properly randomized networks \cite{Guimera2} ($p$-$value<10^{-6}$), while the BC model produced good approximations to the empirical modularity (see Table 1). Shifting our focus to the individual level, nodes have different connectivity roles according to their number and distribution of connections within and outside their own community \cite{Guimera2}. Guimer\`{a} and Nunes Amaral \cite{Guimera2} have heuristically established seven different roles for nodes: roles 1-4 define different non-hub nodes, roles 5-7 are assigned for different classes of hub nodes, and the higher the role the higher the connectivity of the node with other communities. This categorization scheme classifies nodes according to their normalized within-module degree $z_i$ and a participation coefficient $PC_i$ with the rest of the modules. Figure \ref{fig4}A and \ref{fig4}B show the connectivity-role space generated for a pollination and a NYGI network respectively. Note the similar patterns between the two networks. Plants and designers (black dots) are mainly the nodes acting as hubs (roles 5-6), while animals and contractors (red crosses) show high connectivity among modules (roles 3-4). 

\begin{figure}[t]
\begin{center}
\begin{tabular}{cc}
\includegraphics[width=3.5in]{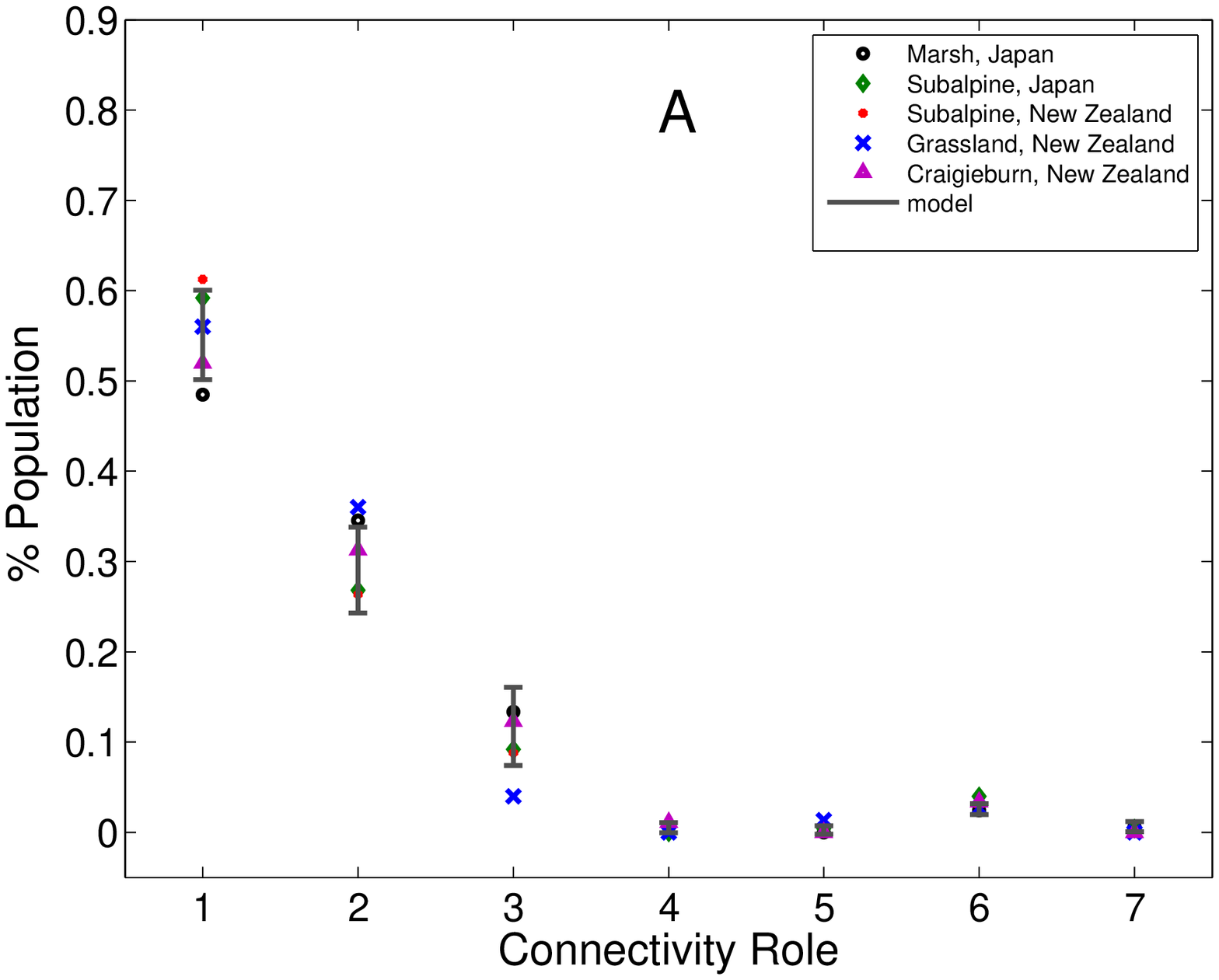}&\includegraphics[width=3.5in]{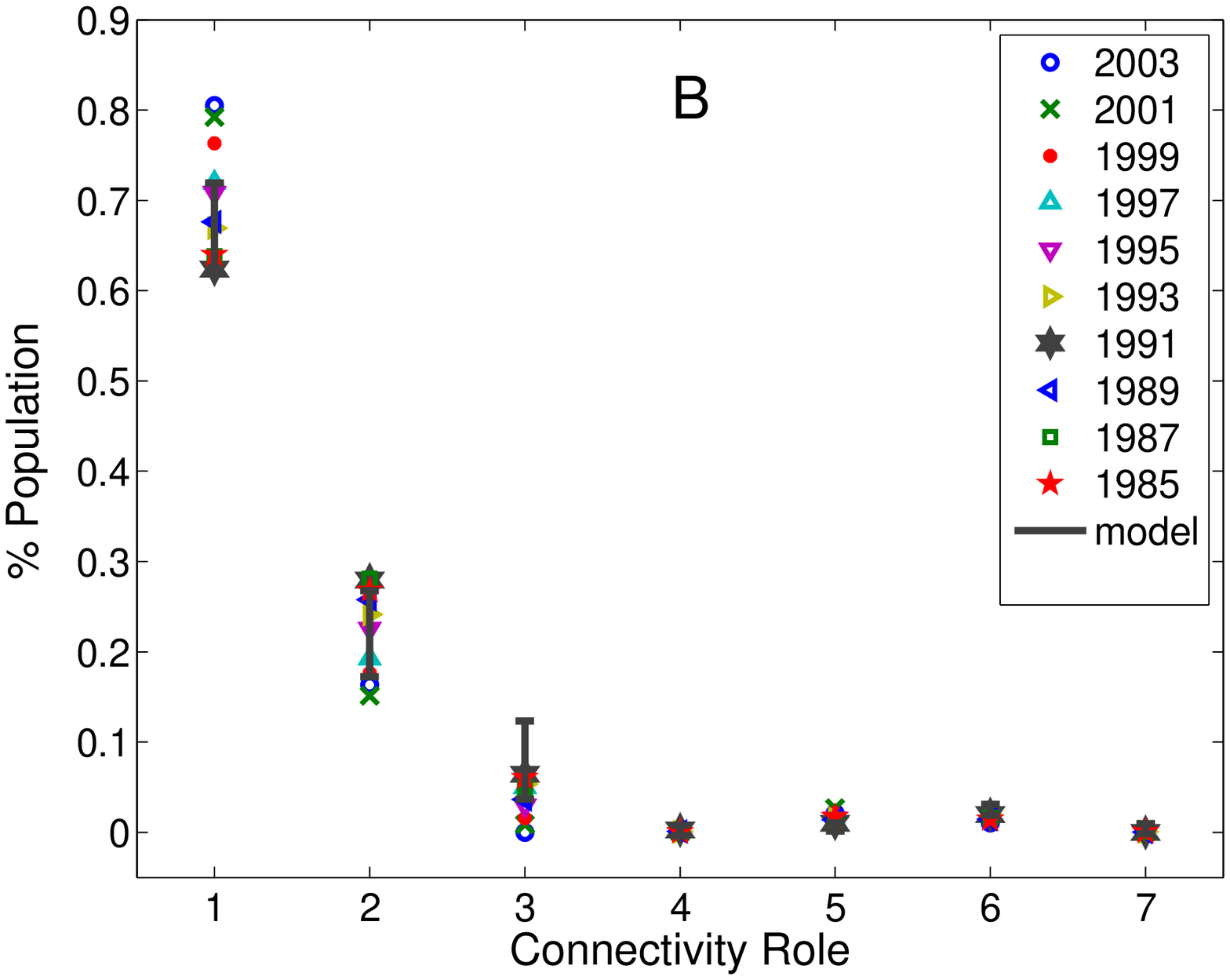}
\end{tabular}
\end{center}
\caption{Model validation for connectivity roles. Panel \textbf{A} and Panel \textbf{B} show the proportion of members within each connectivity role for the modular pollination networks and the NYGI networks respectively. The bars show the accuracy of the BC model to reproduce the same number of nodes within each connectivity role (circles) as the ones observed in the empirical networks. Dots correspond to the empirical values and bars ($avg.\pm1s.d.$) correspond to the model-generated values.}
\label{fig5}
\end{figure}

To test the ability of the BC model to accurately reproduce the same number of nodes within each connectivity role observed in the empirical networks, we measured the Pearson correlation and the ratio of the connectivity role norms $d$ for the observed and model-generated networks. The ratio of the norms $d$ is defined by $d=\left|x\right|/\left|y\right|$, where $\left|x\right|=\sqrt{\sum{_{i=1}^m}x_i^2}$; $\left|y\right|=\sqrt{\sum{_{i=1}^m}y_i^2}$; $m$ is the number of connectivity roles; and $x$ and $y$ are the proportion of nodes within each role $i$ for the empirical and model-generated networks respectively. Note that the ratio measures the relative length between two vectors in an $m$-dimensional space, and values within $0.9<d<1.1$ are fractions with a norm comparable to the empirical data. We consistently find values aligned with the empirical measurements for the pollination and NYGI networks ($r=0.98$, $0.9<d<1.1$). In Figures \ref{fig5}A and \ref{fig5}B, bars confirm the accuracy of the BC model ($avg.\pm1s.d.$) to reproduce the same number of nodes within each connectivity role (circles) for pollination networks and NYGI networks respectively.

\section{Ecological niche markets}
As previously mentioned, a key feature in food-web models is the existence of a hierarchical organization established by niche values. In the BC model we also assumed that members in the population are classified according to a niche or hierarchical value. In pollination networks, these niches or hierarchies can be the result of different morphological, temporal and geographical variables that constrain the amount and identity of partners \cite{Olesen,Guimaraes2,Rezende}. Similarly, firms face interaction barriers according to their niche or position in the market \cite{Carroll,Podolny}. This position can be assessed by the diversity of products and the hierarchical nesting of groups \cite{Moody}. Here, to empirically test this organizational feature on the NYGI network, we analyze the extent to which the emerging communities in the network correspond to actual niche markets. 

\begin{figure}[t]
\begin{center}
\begin{tabular}{cc}
\includegraphics[width=3.5in]{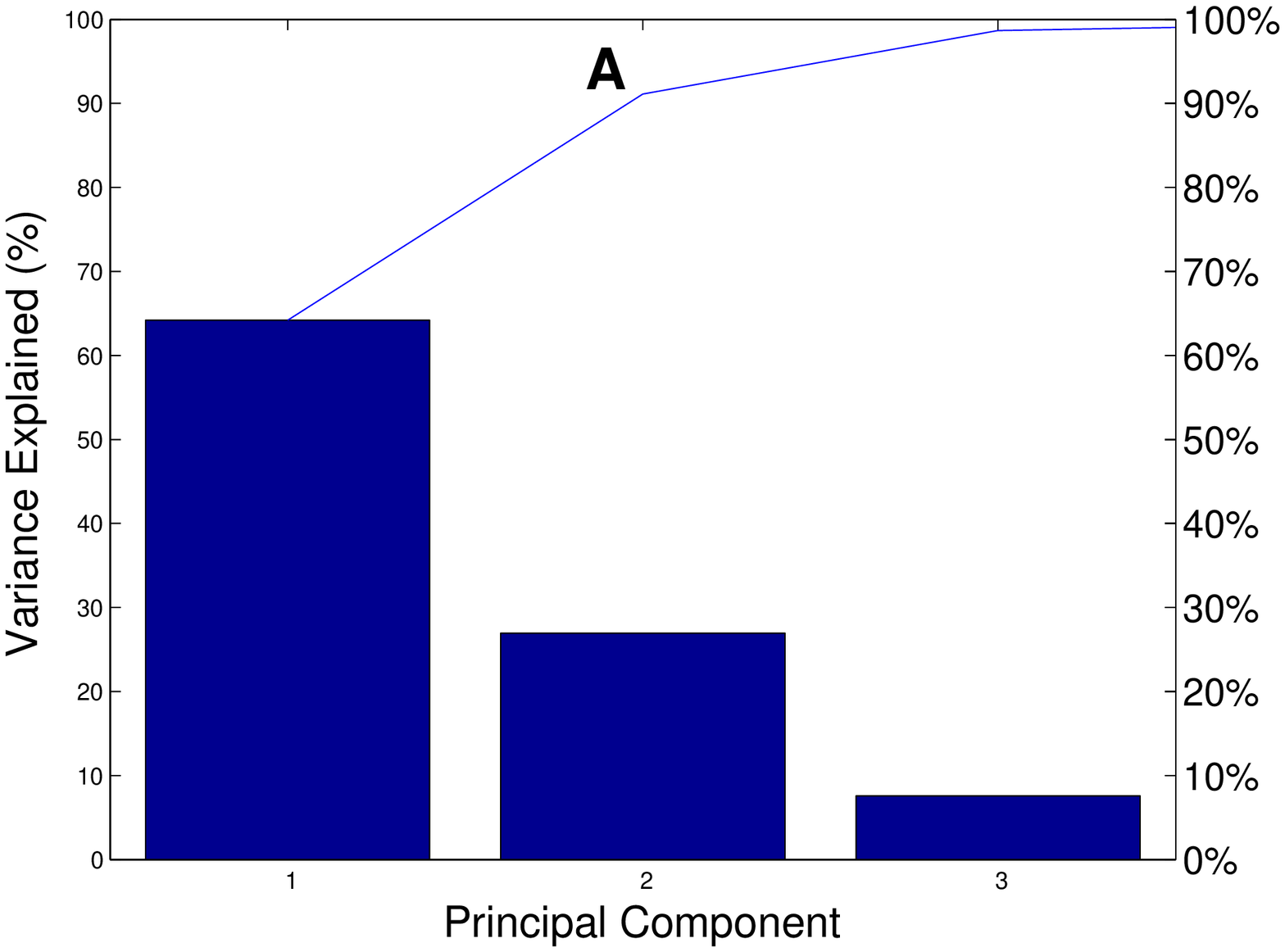}&\includegraphics[width=3.3in]{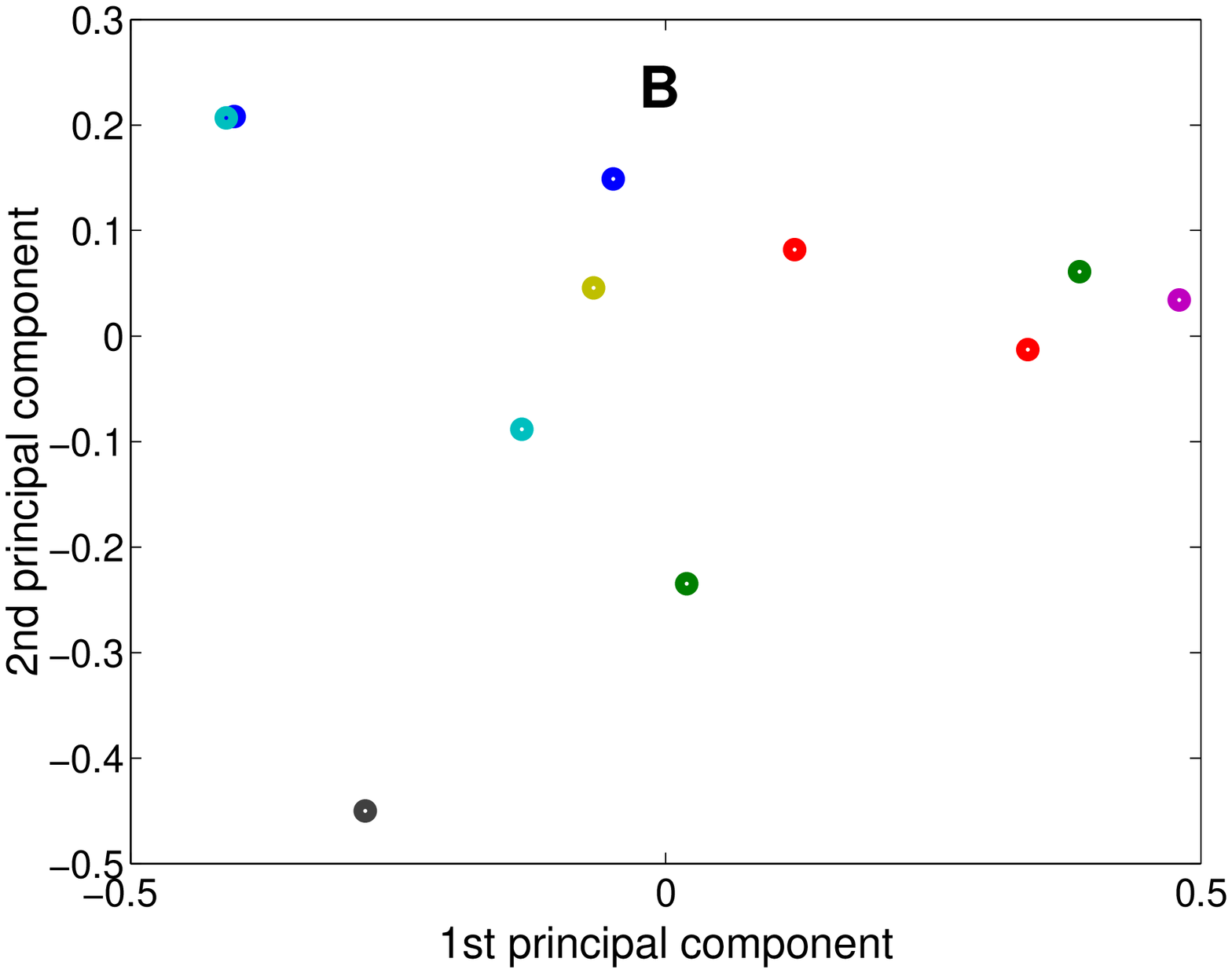}
\end{tabular}
\end{center}
\caption{Niche markets. Panel \textbf{A} shows the variability of each principal component measured for the product correlation $j$ between communities $i$ in the 1990 $P_{ij}$ network. Note that we only need the first two components in order to account for the 95\% of the variability. Panel \textbf{B} shows the transformation of the original $P_{ij}$ matrix by its projection into eigenvectors using the first two principal components. Note that each community $i$ (dots) is defined by a different trend (i.e. position in the two-dimensional space). Different years produce similar results.}
\label{fig6}
\end{figure}

To identify the communities or groups in the network, we use the same community detection algorithm that we used in Section 6.3 to validate the BC model \cite{Guimera2}. To identify the niche of a community, we use data on the different types of garment (i.e. men's coats, women's skirts, t-shirts, etc.) that a firm designs or manufactures. Thus, the diversity of products or garment could be taken as a proxy for the niche characteristics of a particular community. This is to say, two different communities of firms would be in separated niches if they produce different products. We measure the correlation between communities and products using a principal components analysis. To do this we built a matrix $P$, where each element $P_{ij}$ corresponds to the percentage of firms in community $i$ working on garment $j$. Using this matrix, we calculated the correlation of products between communities given by a transformation of values $P_{ij}$ into principal components (i.e. eigenvectors). Using the two largest eigenvalues, which accounted for the 95\% of the variability (see Figure \ref{fig6}A), we generated a two-dimensional projection of the values. Figure \ref{fig6}B shows the projection (dots) and in fact reveals different production trends (i.e different positions in the two-dimensional space) between communities.

To test whether the niche or product differentiation is an artifact of the different number of firms in a community, for each community $i$ we measure its level of production diversification defined by $d(i)=\sum_j (P_{ij})^2$. Here, $d(i)=0$ if every firm in the community works on a different product, and $d(i)=1$ if all firms work on a single type of product. The correlation between the size of a community and its level of product diversification was $r=0.01$, which confirms that the network's organizational structure is confirmed by non-structural niche barriers. Hence, we believe future studies should investigate how the model-generated ordering map into empirical hierarchies at the level of individuals.

\section{Organizing mechanisms}
Social networks, metabolic networks, socio-economic networks, and ecological networks, among others display network structures that differ from random assemblages \cite{Newman1,Dorogov,Caldarelli,Saavedra1}. These comparisons are based on an appropriate randomization of the network (i.e. null model), which assigns a probability $p$ to the existence of each pairwise interaction. These null models can range from models that only preserve the number of connections in the network (e.g. $p=L/S^2$) to models that keep the actual distribution of connections intact and allow a further exploration of the structural correlations in the network \cite{Watts,Maslov}. However, null models based on organizing mechanisms rather than structural characteristics in the network have not been fully explored. Here, we make use of the BC model to show the effects that different constraints acting over the organizing mechanisms of mutualistic networks might produce on the hierarchical or nested structure of ecological networks, which is a key feature in the robustness and biodiversity of these systems \cite{Bascompte3,Bastolla}.

\begin{figure}[t]
\begin{center}
\begin{tabular}{c}
\includegraphics[width=4in]{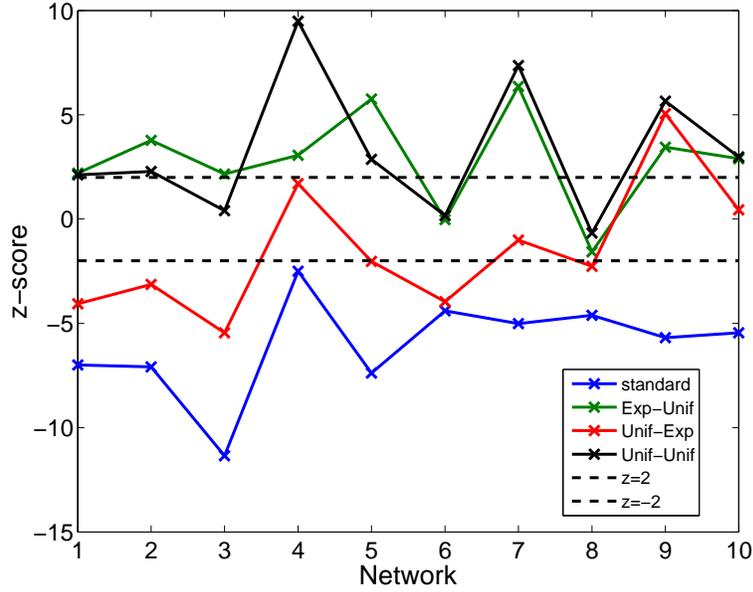}
\end{tabular}
\end{center}
\caption{An alternative null model. The figure shows the $z$-score values (y-axis) for each of the pollination networks (x-axis). The $z$-score values are given by $Z=(N-\overline{N}_m)/\sigma_m$ where $N$ is the nestedness value for the empirical network, and $\overline{N}_m$ and $\sigma_m$ are the average nestedness value and the standard deviation for the null-model-generated network respectively. The blue line and red line correspond to the standard null model II \cite{Bascompte3} and the modified version of the BC model using a uniform distribution for the specialization factor $\lambda_p$ respectively. Note that both null models produce a majority of $Z<-2$, which indicates that the empirical values are significantly nested compared to these null models. In contrast, the green line and black line correspond to the modified versions of the BC model using a uniform distribution for the interaction factor $\lambda_{lp}$ alone and using a uniform distribution for both factors respectively. Note that both null models produce a majority of $Z>2$, which indicates that the empirical values are significantly less nested compared to these null models. Dashed lines are just eye guidelines for $Z=2$ and $Z=-2$, where values above or below these lines correspond to statistical deviations from empirical values.}
\label{fig7}
\end{figure}

In the BC model, the number and identity of partners are constrained by a hierarchical ordering, which are modulated by an external factor $\lambda_p$ and $\lambda_{l_{pi}}$ acting over the specialization and interaction mechanisms respectively. In line with food web models \cite{Stouffer}, we have assumed random inhomogeneous values for $\lambda_p$ and $\lambda_{l_{pi}}$ given  by an exponential distribution. In food webs, this distribution imposes a specific range of possible interactions, which might correspond to the actual handling and foraging capacities of species \cite{Petchey}. To study the effects of these constraints on the specialization and interaction mechanisms of the BC model, we replaced the exponential distribution by a randomly uniformed distribution. By modifying the specialization mechanism alone (i.e. $\lambda_p$), we found nestedness values below the ones displayed by the observed pollination networks. In line with standard null models, Figure \ref{fig7} shows an under-representation of nestedness by using both the modified version of the BC model (red line) and the standard null model II (blue line) \cite{Bascompte3}. This suggests that the exponential distribution limiting the number of connections for plants might in fact be responsible for the nested organization of pollination networks. However, if we modify the interaction mechanism (i.e. $\lambda_{l_{pi}}$), we found the opposite behavior. Figure \ref{fig7} (green line) shows an over-representation of nestedness compared to the observed networks. Figure \ref{fig7} (black line) confirms that this pattern is robust even if we modify both the specialization and the interaction mechanisms simultaneously. These results suggest that the organization of mutualistic networks is neither extremely nested nor random, which might emerge as a compromise between the structural robustness (specialization mechanism) and the functionality of the network (interaction mechanism). We believe this conjecture should be tested in future work. 

\section{Conclusions}
The study of direct member-to-member interactions have allowed us to find that the structure of ecological and socio-economic networks generated by mutually-beneficial interactions exhibits remarkably similar features. This empirical finding motivates the proposed model for bipartite cooperation, starting from a generalization of the niche model \cite{Williams}, which can successfully reproduce the overall structure of pollination and NYGI networks using the number of members and the total number of links as the only input parameters. We have identified common organizing mechanisms operating in these radically different networks, which are the result of a hierarchical ordering that favors the presence of asymmetric interactions between their members. These organizing mechanisms are exponentially constrained by external factors (such as environmental and market pressures), which modulate the number and identity of potential partners. Using our model, we investigated the effects that different constraints acting on the specialization and interaction mechanisms of mutualistic networks might produce on the hierarchical arrangement of these networks. We have found that real-world mutualistic networks display network organizations, which are between highly nested and random structures. This could enable a systematic approach to study the interplay between functionality and structural robustness in mutualistic networks. The success of this simple stochastic model in generating the overall structural characteristics of mutualistic networks makes it a suitable starting point for more elaborate ecological and socio-economic models which seek to understand the effects of changes in population size, number of interactions, and harsher environmental or market conditions.

\newpage

\begin{table}[h]
	\scriptsize
	\centering
		\begin{tabular}{ccccccc}
			Dataset-environment&\textit{L}&\textit{P}&\textit{A}&$KS_P-KS_A$&$N$&$Q$ \\ [2mm]  \hline
			Kato and Miura (1996)&430&64&187&0.326$\dag\dag-0.438\dag\dag$&0.976$\dag\dag$(0.969)&0.551$\dag\dag$(0.553)\\ [2mm] 
			Primack (1983)&374&41&139&0.633$\dag\dag-0.385\dag\dag$&0.957$\dag\dag$(0.96)&0.474$\dag$(0.465)\\ [2mm]  
			Kato et al. (1993)&865&90&354&0.552\dag\dag-0.001*&0.985$\dag$(0.976)&0.545$\dag$(0.532)\\ [2mm] 
			Primack (1983)&120&18&60&0.108$\dag-0.999\dag\dag$&0.858*(0.936)&0.553$**$(0.527)\\ [2mm] 
			Primack (1983)&346&49&118&0.002*-0.001*&0.961$\dag\dag$(0.955)&0.480$\dag$(0.468)\\ [2mm]  
			Hocking (1968)&179&28&81&0.097$\dag-0.989\dag\dag$&0.971$\dag$(0.950)&nm\\ [2mm] 
			Inouye and Pyke (1988)&252&36&81&0.608$\dag\dag-0.076\dag$&0.935$\dag\dag$(0.949)&nm\\ [2mm] 
			Schemske et al. (1978)&65&7&33&0.911$\dag\dag-0.642\dag\dag$&0.953$\dag$(0.930)&nm\\ [2mm] 
			Elberling and Olsesen (1999)&453&31&75&0.038**$-0.118\dag$)&0.793*(0.914)&nm\\ [2mm] 
			Elberling and Olsesen (1999)&242&24&118&0.223\dag-0.005**)&0.927$\dag$(0.952)&nm\\ [2mm] 
			NYGI (1985)&7250&823&2562&0.061$\dag-0.115\dag$&0.997$\dag$(0.996)&0.598*(0.502)\\ [2mm] 
			NYGI (1991)&3981&325&1590&0.101$\dag-0.531\dag\dag$&0.994$\dag\dag$(0.993)&0.601*(0.529)\\ [2mm] 
			NYGI (1997)&1450&148&700&0.003**$-0.264\dag$&0.990$\dag$(0.988)&0.653**(0.625)\\ [2mm] 
			NYGI (2003)&228&62&128&0.370$\dag-0.002$**&0.976$\dag$(0.969)&0.711$\dag$(0.700)\\ \hline
		\end{tabular}
		\caption{Model Validation.  For ten pollination datasets and four NYGI networks used in this paper, the table presents its source; total number of links $L$, $P$ and $A$ are the number of members in class P and class A respectively. For the degree distributions, ($KS_P-KS_A$) shows the combined Kolmogorov-Smirnov (KS) probability using the two-group equivalence KS test between the empirical and model-generated distributions for class P and class A respectively. $N$ and $Q$ correspond to the observed nestedness and mean modularity values respectively, along with the normalized errors ($z$-score) for the comparison between the empirical and model-generated values. The model-generated mean values for $N$ and $Q$ are shown inside the parentheses. Five of the observed pollination networks have already been found to be non-modular (nm) \cite{Bascompte2}.		
		 \\$\dag\dag$: $KS>0.30$ or normalized errors $<1$ model s.d. (excellent fit); $\dag$: $KS<0.30$ or normalized errors between 1 and 2 model s.d. (good fit); **: $KS<0.05$ or normalized errors between 2 and 3 model s.d. (poor fit); *: $KS<0.01$ or normalized errors $>3$ model s.d. (bad fit).}
	\label{table1}
\end{table}

\newpage

\end{document}